\documentstyle[12pt,epsfig]{article} 
\begin{document}
\baselineskip=0.7cm

\begin{flushright} Preprint USM-TH-73 \end{flushright}
\vspace*{.2in}

\begin{center} \Large\bf Neutrino Scattering Off Spin-Polarized Particles 
in Supernovae      \end{center}
\vspace*{.10in}

\begin{center}
P.D. Morley\footnote[1]{morleyp@cna.org}  
\\CNA Corporation \\ 4401 Ford Avenue\\Alexandria, VA 22302-1498 \end{center}
\vspace*{.10in}
\begin{center}
C. Dib\footnote[2]{cdib@fis.utfsm.cl} and 
I. Schmidt\footnote[3]{ischmidt@fis.utfsm.cl} 
\\ Department of Physics \\ Universidad T\'{e}cnica 
Federico Santa Mar\'\i a \\
Casilla 110-V \\ Valpara\'\i so, Chile  
\end{center}
\vspace*{.10in}
\begin{abstract}
\baselineskip=0.7cm
We calculate neutrino and anti-neutrino scattering off electrons and 
nucleons in supernovae using a detailed Monte Carlo transport code 
incorporating realistic equations of state. The goal is to 
determine whether particles in a neutron star core, partially spin-aligned by 
the local magnetic field, 
can give rise to asymmetric neutrino emission via the standard parity 
breaking weak force. We conclude that electron scattering in a 
very high magnetic field does indeed give a net asymmetry, but 
that nucleon scattering, if present, removes the
asymmetry. \end{abstract}

\vspace{.15in}

\newpage
\parskip=9pt

PSR 2224+65 has a measured \cite{Cordes} transverse speed $\ge 
800$ km/s, indicating that it has enough kinetic energy to leave 
the gravitational potential well of our galaxy, which requires a 
speed of 450 km/s \cite{Allen}. It now appears that many 
pulsars have abnormally high proper speeds \cite{Harrison}. Since
neutrino emission carries off 99\% of the binding energy of the 
magnetized neutron star/pulsar formed in supernova explosions, it
is particularly attractive to consider models of asymmetric 
neutrino core emission as the source of the pulsar kick, since 
then only a small fraction of $O$(1\%) of that energy is required
to be asymmetrically radiated. In this regard, several authors 
have proposed mechanisms to accomplish this. Bisnovatyi-Kogan 
\cite{Kogan} considers toroidal magnetic field configurations in 
a differentially rotating star. In the absence of dissipative 
processes, the neutron star returns to the state of rigid 
rotation by loosening the induced higher moment fields via 
magnetorotational explosion. However, it is not very clear how this 
gives rise to asymmetric neutrino luminosities. Kusenko and 
Segr\`{e} \cite{Kus2}, using the approximation of the last scattering surface 
method, propose that neutrino oscillations induced
by non-zero neutrino masses alter the shape of the neutrinosphere, and derive 
a necessary tau neutrino mass of $\sim 100$ eV. 
Grasso et al \cite{Grasso} propose neutrino asymmetries without 
neutrino mass requirements, based on weak violation of 
universality.
We want to study the possibility of explaining the effect within the Standard Model, 
without relying on last scattering assumptions. Instead, we simulate the 
complete process of neutrino birth and transport within the core. We find that 
the last scattering surface assumption does not reproduce the neutrino 
transport properties. While this paper was being prepared, 
Lai and Qian \cite{Lai} have considered the possibility that spin
aligned nucleons could lead to asymmetric neutrino emission. We, 
in fact, find just the opposite: nucleons are parasitic and it is
the electrons which have the potential to lead to asymmetric 
emission.

The spin-aligned fraction, 
$\eta \equiv [N(+) - N(-)]/[N(+) + N(-)]$,
for non-relativistic non-degenerate particles (i.~e.~nucleons in 
proto - neutron star cores) with magnetic moment $\mu_{M}$ in a 
magnetic field $B$ at temperature $T$ is 
\begin{equation}
\eta = \tanh \frac{\mu_{M} B}{kT}, \end{equation}
while for 
relativistic ideal degenerate particles (i.\ e. electrons in neutron star cores) 
(with $T << E_F$, $E_F=$ Fermi energy), 
\begin{equation}
\eta = \frac{(E_F +
\mu_{M}B)^{3}-(E_F- \mu_{M}B)^{3} }{(E_F + \mu_{M}B)^{3}
+ (E_F - \mu_{M}B)^{3}}.
\end{equation}
A cooling proto-neutron star core at $T$ = 5 MeV, $B$ = $10^{14}$ Gauss, 
with baryon density of $ 5 \times 10^{38}$ baryons/cm$^{3}$, 3\% of it as 
protons, has 
$\eta_{\rm neutron} = 1.2 \times 10^{-4}$ and
$\eta_{\rm electron}= .011$. Thus nucleons are 
hardly aligned at all, and
electrons, due to the relativistic degeneracy, have also a rather small 
spin alignment. Only for gigantic fields $B \sim 2000$ GT will the ideal gas 
electrons be highly spin aligned in neutron stars. It is possible, however,
that electrons do not form an ideal relativistic degenerate gas within neutron 
stars, but instead become nearly a perfect ferromagnetic spin aligned fluid, 
with a nearest neighbor Hamiltonian 
$ \sim \; \sum_{i,j} \vec{s}_{i} \cdot \vec{s}_{j} h_{i,j}$ with $h_{i,j}<0$. 
This may happen since parallel spin alignments will tend to reduce the Coulomb 
repulsion between electrons, a major energy consideration at these high number 
densities. If such an electron ferromagnetic fluid were present, then electron 
spin alignment would be nearly 100\%
even for $10^{14}$ Gauss. In our simulation, we just use an ideal electron gas, 
disregarding possible ferromagnetic effects.

There has been some recent discussion in the literature 
indicating that no asymmetry can arise from neutrino multiple 
scattering if the neutrinos are in {\sl thermal equilibrium} 
\cite{Kusenko}. We fully agree with this statement, although it 
is not the case we are considering here. The neutrinos that 
produce the asymmetry do not reach thermal equilibrium, due to a non-negligible
neutrino mean free path comparable to the temperature gradient scale. In fact, 
too little or too many collisions, leads to zero neutrino asymmetry, but 
we find that a moderate number of collisions per neutrino of $\sim 4000$ 
leads to the effects reported here.

The probability $P(\theta_{4},\phi_{4})d \theta_{4} d \phi_{4}$ that a 
neutrino or anti-neutrino
will scatter into the solid angle $d \Omega_{4}=
\sin \theta_{4} d\theta_{4} d\phi_{4}$ is
      \begin{equation} P(\theta_{4},\phi_{4}) d\theta_{4} d\phi_{4} = 
      \frac{|M(p_{1},p_{2},p_{3},p_{4})|^{2} sin \theta_{4}}
      {\int |M(p_{1},p_{2},p_{3},p_{4})|^{2}d\Omega_{4}} d\theta_{4} 
      d\phi_{4} \; ,
      \end{equation}
where $M(p_{1},p_{2},p_{3},p_{4})$ is the scattering amplitude 
for a (anti) neutrino with incoming momentum $p_2$ and outgoing 
momentum $p_4$, off a target with initial and final momenta $p_1$
and $p_3$ respectively.
For scattering of neutrinos off electrons, the spin-dependent 
matrix element squared is
  \begin{eqnarray} |M(e_{1} \nu_{2} \rightarrow e_{3} \nu_{4})|^{2}  
  =  16G_{F}^{2}
  \{L_e ^2[(p_{4} \cdot p_{3})(p_{2} \cdot p_{1}) - m_{e} (p_{4} 
\cdot   p_{3})(p_{2} \cdot s_{1})] & \mbox{} \nonumber \\
  + R_e ^{2}[(p_{4} \cdot p_{1})(p_{3} \cdot p_{2}) + m_{e}
  (p_{3} \cdot p_{2})(p_{4} \cdot s_{1})] & \mbox{} \\
  -L_e R_e [m_{e}^{2}(p_{4} \cdot
  p_{2})+m_{e}(p_{2} \cdot p_{1})(p_{4} \cdot s_{1}) - m_{e}(p_{4} \cdot
  p_{1})(p_{2} \cdot s_{1})] \} & \mbox{} , \nonumber
  \end{eqnarray}
where the weak couplings $L_e$ and $R_e$ are shown in Table~1, 
and $s_1$ is the polarization four-vector of the initial electron, which 
satisfies $p_1\cdot s_1 =0$ and $s_1 = -1$. For antineutrino scattering, 
the respective expression is the same as the 
one above after exchanging $p_2 \leftrightarrow p_4$.

  \begin{table} 
     \begin{tabular}{|l|cc||l|cc|}  \hline  
     \mbox{} &  $n$-$\nu$ & $p$-$\nu$ & \mbox{} & $e$-$\nu_e$ & $e$-$\nu_{\mu,\tau}$ \\ \hline  
     $c_v$ & $-1/2$ & $(1-4\sin^2 \theta_W )/2$ & $L_e$ & $2\sin^2 \theta_W +1$ 
& $2\sin^2 \theta_W - 1$ \\
     $c_a$ & $-g_A/2$ & $g_A/2$ & $R_e$ & $2 \sin^2\theta_W$ & $2\sin^2\theta_W$ \\ \hline
     \end{tabular}
     \caption{The couplings $c_v$ and $c_a$ for nucleon scattering (Eq.~7), 
and $L_e$ and $R_e$ for electron scattering (Eqs.~4-5). 
For nucleon scattering with antineutrinos, $c_a$ changes sign; for electron
scattering with neutrinos or antineutrinos the couplings are the same. We use 
$\sin^2\theta_W = 0.23$ and $g_A = 1.26$.}
     \end{table}

For nucleon scattering, Ref.~\cite{Horowitz} has given the scattering probability
    \begin{equation} P(\theta_{4},\phi_{4})d \theta_{4} 
    d\theta_{4} d\phi_{4} = \frac{
    \frac{d \sigma}{d \Omega_{4}} 
    \sin \theta_{4} }{ \int{ d \sigma}} 
    \; d \theta_{4} d \phi_{4},
    \end{equation}
where the cross section for neutrino-nucleon 
scattering is:     
\begin{equation} \frac{d \sigma}{d \Omega_{4}}
=\frac{G_F ^{2}E_\nu ^{2}}{4\pi^{2}}\{ c_{v}^{2}+3 c_{a}^{2}+(c_{v}^{2}-c_{a}^{2})
    \cos \theta
+ 2 \eta c_{a} [(c_{v}-c_{a})  
    \cos \theta_{out}
 +(c_{v}+c_{a})\cos \theta_{in} ]  \}
\end{equation}
Here $\eta$ is the polarization factor of Eq.~(1), $c_v$ and $c_a$ are given 
in Table~1, $\theta$ is the scattering angle, and 
$\theta_{in}$ and $\theta_{out}$ are the angles between the nucleon spin and the 
incoming or outgoing neutrino momentum, respectively. 
In Table~2 we give the star parameter ranges used in our simulations, 
including the assumed large magnetic fields associated with magnestars.      
  \begin{table} 
     \begin{tabular}{|cl|}  \hline  
     \multicolumn{2}{|c|}{Star Parameters} \\ \hline  
     \hbox{\hskip 0.3in $B$ \hskip 0.3in}& $0.1 \rightarrow 5000$  [GTesla] \\
     $T$  & $1 \rightarrow 100$ [MeV] \\  
     $n_{B}$  & $ 5 \times
     10^{37} \rightarrow 1.16 \times 10^{39}$ [baryon/cm$^{3}$] \\
     $f_{e}$ & $0.001 \rightarrow 0.5$  \\ \hline
     \end{tabular}
     \caption{The physical parameters involved in the asymmetry calculation.}
     \end{table}

We now add the astrophysics. The density of matter in the interior of neutron 
stars increases from $10^{25}$ baryon/cm$^{3}$ in the surface layer to 
$\sim 10^{39}$ baryon/cm$^{3}$ near the center. However, more than 95\% 
of the mass of a typical neutron star is compressed to baryon densities 
exceeding $10^{38}$ cm$^{-3}$ \cite{Bowers} and, 
to an excellent approximation, the supernova core can be described as 
hadronic matter at constant density. The different equations 
of state (EOS) give rise to different
fractions of electrons present, $f_{e}=n_{e}/n_{B}$, where 
$n_{e}$ and $n_{B}$ are the electron and baryon densities
respectively. Thus we describe the EOS landscape by two parameters: 
$n_{B}$ and $f_{e}$. We set the mass of the core to 1.8 $M_{\odot}$. 
The thermodynamic parameters are the 
temperature $T$ and the interior magnetic field. We 
use constant interior magnetic fields. With respect to neutrino 
production, we only consider neutrino-anti-neutrino thermal 
pairs, as 90\% of the neutrino emission is of thermal origin.

A Monte Carlo neutrino transport code for neutrino scattering off
polarized targets (electrons only or both electrons and nucleons)
was written to determine if the parity violation in the weak 
force can lead to asymmetric neutrino emission from the surface, and thus possibly 
explain the large proper motion of pulsars. Because of 
Fermi degeneracy, essentially only those electrons near the Fermi level 
participate in scattering. Since the electrons do not interact via the hadronic 
force (and the star is electrically neutral) they might be described as an ideal 
relativistic degenerate Fermi gas with Fermi momentum 
$p_{F}=6.10464 \times 10^{-11}(f_{e}n_{B})^{1/3}$ MeV/c, with $n_B$ 
in $1/{\rm cm}^3$
(but see above qualifications concerning possible electron ferromagnetism).
 An experimental constraint on the EOS is that the radius of the 
neutron star must be smaller than 14 - 16 km \cite{Walter}. 

The simulation goes as follows: a random flavor $\nu$-$\bar\nu$ pair is born 
at a random location within the core, with a spherically random direction
and each particle with an energy equal to 3$T$, where $T$ is the temperature 
at that position in the star. A discrete temperature 
gradient was imposed. For this purpose, four concentric shells of
inner and outer radiae ${n\over 4} R_{\rm star}$ and ${n+1\over 4} R_{\rm star}$ 
respectively ($n=0,1,2,3$) were considered. The temperature in each shell 
was set to ${6-n\over 3} T_s$, where $T_s$ is a temperature scale parameter. 
Thus for birth location in the $n$th shell the neutrino energy was 
$E_{\nu,\bar{\nu}}=(6-n) T_{s}$.
For a uniform temperature in a constant density core, the $\nu$-$\bar\nu$ 
pair emission probability per unit volume is also uniform; the production 
point thus has a radius $r \sim R_{star}\times\xi^{1/3}$, where 
$\xi \in$ [0,1] is a uniformly distributed random variable. Instead, 
with negative temperature gradients (hotter center region), the emission is 
also higher at the center, skewing the distribution towards 
$r \sim R_{star}\times\xi$. We used here 
$r = 0.4 R_{\rm star} \times \xi$, with the phenomenological 
factor 0.4 to take into account $\nu$-$\bar\nu$ production only within the 
interior core. After birth, the location of the next scattering event is 
statistically generated from the 
distribution given by the mean free path $\ell = 1/(\sum_i n_i \sigma_i)$ ($n_i$ 
is the $i$-type target density --electrons or nucleons-- and $\sigma_i$ 
the cross section)
\cite{explanation}. If this is outside the radius, the neutrino 
energy and momentum components are accrued as ejected flux. 
If the scattering position is still within the core, a randomly directed 
electron (in the case of electron collisions) from the Fermi surface is picked, 
with its spin direction parallel or anti-parallel to the magnetic field, 
determined from the spin probability distribution function. The probability of 
parallel/antiparallel spin is defined in terms of $\eta$ from Eq.~(2) as
$P_{\pm}=(1 \pm \eta)/2$. 
The direction of the outgoing neutrino is then generated according to the 
distribution given by the differential cross section (Eq.~3) and a Pauli check 
is made. 
Then the outgoing neutrino seeks the next scattering event. 
This is continued until the
neutrino leaves the star. Finally, the history of the neutrino's birth 
partner is followed. This is done for $N$ neutrinos.

The asymmetry is defined as the accrued ejected
momentum over the accrued ejected energy:
\begin{equation}
Asymmetry = \frac{\sqrt{[\sum p^x]^2 + [\sum p^y]^2 + [\sum p^z]^2}} {\sum E_i}
\Longrightarrow
\frac{|\sum p^z|}{\sum E_i},
\end{equation}
where $z$ is the direction of the magnetic field\cite{explanation2}, 
We verified that the other momentum 
components, $\sum p^x$ and $\sum p^y$, fall within the statistical 
fluctuations, while $\sum p^z$ 
stands out as the signal. The variance of the statistical 
fluctuations increases as $\sim \sqrt{N}$, so the 
asymmetrical statistical `background' decreases $\sim 1/ \sqrt{N}$. 
To obtain a $O$(1\%) positive signal with a $\sim$ .1\% error thus requires $N 
\sim 10^{6}$. In the actual runs, one million pairs were followed, giving
$N=2 \times 10^{6}$. The `background' statistical asymmetry was calculated by
performing a run with random spin orientation at each scattering
location, which, for the samples of a million pairs gave a typical magnitude
of 0.05\%. For runs including both electron and nucleon scattering, the 
probability of electron scattering is 
$P_{\rm electron} = \Sigma_{\rm electron}/(\Sigma_{
\rm electron}+ \Sigma_{\rm proton} + \Sigma_{\rm neutron})$,
where $\Sigma_{i} = n_{i}$ (particle density) $ \times  
\sigma_{i}$ (scattering cross section). Similar probabilities exist for 
scattering off neutrons and protons. 

The results we present in our tables and figures correspond to a neutron star 
with 1.8 solar masses, radius 8.3 km and electron/nucleon ratio $f_{e} = 0.12$. 
This size and structure is representative of most realistic equations of state 
which incorporate nuclear repulsion at short distances. Typically, the heaviest 
neutron star for such equations of state is over two solar masses 
(see {\it e.g.} R. Bowers {\it et al.} in Ref [10]).

Several critical details arose. Firstly, it is a highly 
non-trivial problem in computational physics to draw scattering 
samples from Eqs.~(3--4). We use the `rejection method' 
\cite{Spanier}, generalized to two random variables. Secondly, 
due to the billions of random numbers used, we discovered that a 
popular random number generator, Knuth's algorithm \cite{Knuth}, 
performed unsatisfactorily. We finally used the `gold-plated' 
random number generator ran2, of Ref.~\cite{Press}, which 
performed without problems.

In Fig.~1 we display the results of a positive signal, corresponding to the 
case of extremely high magnetic field ($\sim 10^3$ GT) and scattering with 
the electron gas only. 
Subtle physics effects are at play here. 
First, looking at Eqs.~(3--4), we see that the spin terms are
multiplied by $m_{e}$, the electron mass; thus, it would seem (wrongly)
that relativistic electrons have little parity violation. Secondly, from 
Fig.~1, the asymmetry increases with temperature, reaching O(1\%) at 
$\sim 17$ MeV. These two observations are related: the magnitude 
of the polarization four-vector for electrons traveling 
parallel or anti-parallel to the magnetic field is proportional to the
relativistic $\gamma$ factor and so for such electrons $m_{e}s \sim E$; 
as the temperature increases, the Pauli window increases allowing a greater 
energy range of allowed transitions. Thus the asymmetry is expected to be an 
increasing function of temperature.

     \begin{table} 
     \begin{tabular}{|ccc|c|} \hline   
     T$_{S}$ (MeV) & B (GT) & target particles & Asymmetry \\ \hline
     10 & 200 & electrons & $(0.18 \pm 0.05)\%$ \\
     10 & 4000 & electrons & $(0.47 \pm 0.05)\%$ \\
     10 & 4000 & nucleons \& electrons & $(0.04 \pm 0.05)\%$ \\ \hline
     \end{tabular}
     \caption{Asymmetry in neutrino flux for different magnetic field and 
scattering targets, for a neutron star of 1.8 $M_{\odot}$, radius 8.3 km, 
and electron/nucleon fraction 0.12}
     \end{table}

Now, if we include neutrino scattering with nucleons as well as with electrons, 
the results change dramatically. 
Elastic scattering of neutrinos off hadrons occurs via the neutral current. 
The small magnetic moment of the nucleons gives very little polarization, so 
we would expect that allowance of nucleon scattering would tend to remove the 
neutrino emission asymmetry. Moreover, the nucleon density is much larger than 
the electron density, so that nucleon scattering and its effects should 
dominate. 
In Table~3, we present the results of the nucleon scattering inclusion, 
showing that this is indeed the case: even at extremely high magnetic fields, 
the inclusion of nucleon scattering washes out the asymmetry, as shown 
in the last row of Table~3. Here we should recall that the statistical 
fluctuation of the simulations are about 0.05\% in the asymmetry, which 
is already larger than the signal obtained.

Our numerical results support the hypothesis that neutrino scattering off 
spin-polarized electrons is a viable mechanism for achieving high pulsar proper
velocities in the case where extremely high magnetic fields exist in the core,
and whenever scattering with nucleons can be neglected, 
without resorting to any non-standard physics or neutrino oscillation mechanism,
but if nucleon scattering is present, the asymmetry disappears.
Now, since the nucleon density is necessarily larger than the electron density, 
to neglect nucleon scattering in favor of electron scattering does not seem 
plausible, unless there is an additional mechanism, like superfluid phases 
in the core, that could block nucleon scattering. 

\bigskip
\begin{flushleft} {\Large \bf Acknowledgments} \end{flushleft}

We thank Olivier Espinosa for insightful discussions. 
This work was supported in part by Fondecyt (Chile) contracts 1980149 and 
1980150. I.S. also received support from a {\sl C\'atedra Presidencial en 
Ciencias} (Chile).

\vspace{0.5cm}
\begin{figure}[htb]
\begin{center}
\leavevmode {\epsfysize=10cm \epsffile{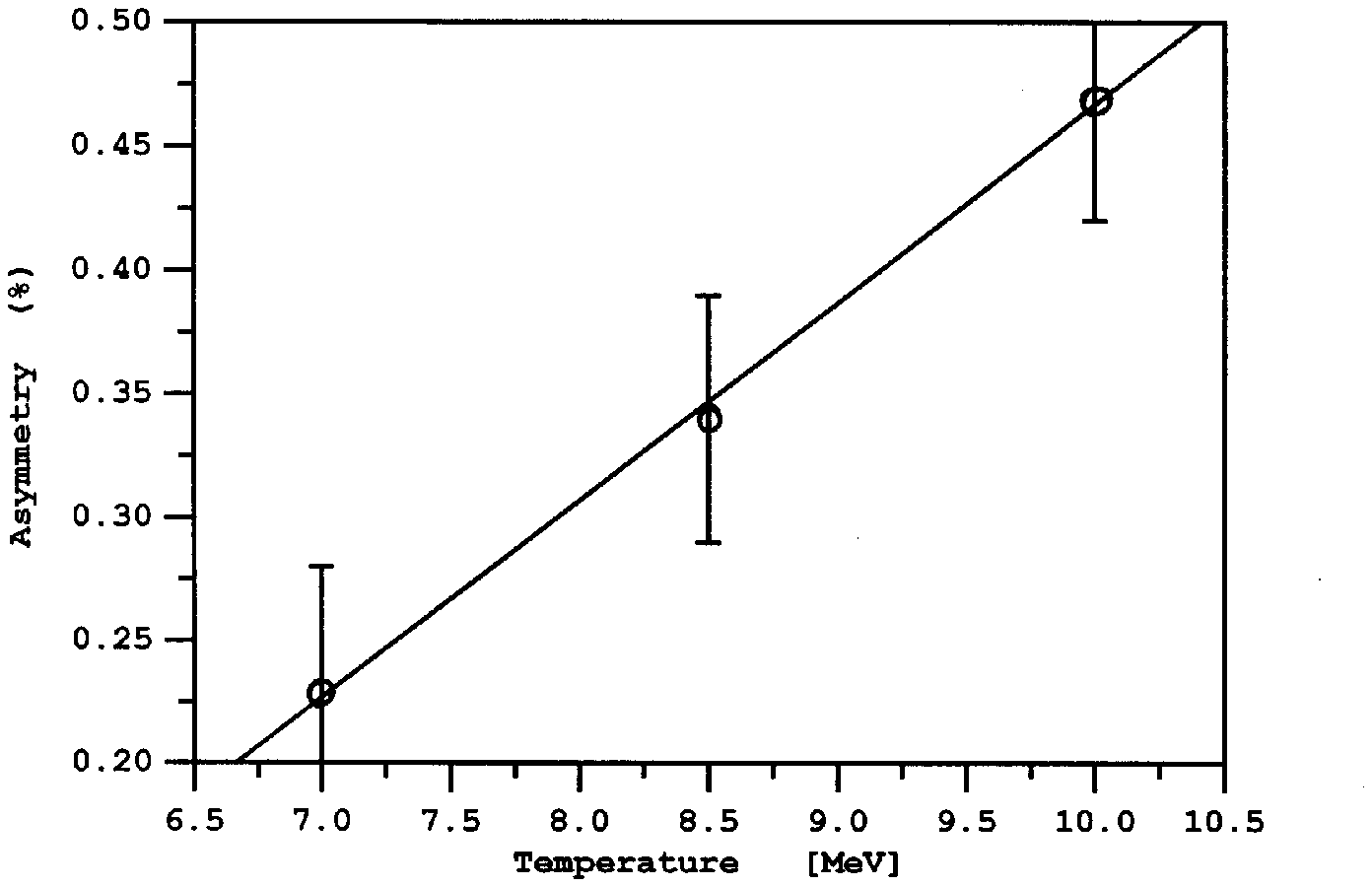}} 
\end{center}
\caption[*]{\baselineskip 13pt 
Asymmetry vs.~Temperature, for a star with baryon density 
$n_B = 8\times 10^{38}~1/{\rm cm}^3$, radius $R_{\rm star} = 8.3$ km, 
electron fraction $f_e$ = 0.12, and magnetic field $B = 4000$ GTesla.}
\label{figure1}
\end{figure}
 

\begin{thebibliography}{99}
\bibitem{explanation2} By symmetry, $z$ is the only preferencial direction.
\end{thebibliography}
\end{document}